\documentclass[letter]{aa} 

\usepackage{graphicx}
\usepackage{txfonts}
\newcommand{\xmm}{{\em XMM-Newton}}

\newcommand{\cxo}{{\em Chandra}}

\newcommand{\eso}{ESO\,338-4}

\newcommand{{\myr}}{\mbox{[$M_\odot\,{\rm yr}^{-1}$}]}
\newcommand{\lsim}{\raisebox{-.4ex}{$\stackrel{<}{\scriptstyle \sim}$}}
\newcommand{\msim}{\raisebox{-.4ex}{$\stackrel{>}{\scriptstyle \sim}$}}

\newcommand{\flux}{erg\,cm$^{-2}$\,s$^{-1}$}
\newcommand{\new}{}
\newcommand{\nw}{}

\usepackage[normalem]{ulem}

\begin{document} 


\title{ULX contribution to stellar feedback: 
an intermediate-mass black hole candidate and the population of ULXs in 
the low-metallicity starburst galaxy  ESO\,338-4\thanks{Based  on observations obtained with the science missions \xmm\ (ObsID 0780790201) and \cxo\ (ObsID 21866).}}
\titlerunning{ULXs in starburst galaxy \eso}
\author{Lidia M.\ Oskinova\inst{1,2}, 
        Arjan Bik \inst{3}, 
        J. Miguel Mas-Hesse \inst{4},
        Matthew Hayes \inst{3},
        Angela  Adamo \inst{3},
        G{\"o}ran {\"O}stlin \inst{3},
        Felix F\"urst \inst{5},
        H\'ector Ot\'{\i}--Floranes \inst{4}}
          
   \institute{\inst{1}{Institute for physics and astronomy, University of 
Potsdam, 
              Karl-Liebknecht-Str. 24/25, D-14476 Potsdam, Germany}\\
              \email{lida@astro.physik.uni-potsdam.de}  \\    
              \inst{2}{Department of Astronomy, Kazan Federal University, Kremlevskaya Str 18, Kazan, Russia} 
\\
              \inst{3}{Department of Astronomy, Stockholm University, Oscar Klein Centre, AlbaNova University Centre, 106 91, 
              Stockholm, Sweden}  
\\              
              \inst{4}{ Centro de Astrobiolog\'ia (CSIC--INTA), 28850 Torrejon de Ardoz, Madrid,Spain}
  \\
\inst{5}{European Space Astronomy Centre (ESAC), Science Operations Departement,
28692 Villanueva de la Ca\~nada, Madrid, Spain}
   \date{Received ? / Accepted ?}
}

 
  \abstract
{X-ray radiation from accreting compact objects is an important part of stellar feedback.  The metal-poor
 galaxy \eso\ has experienced vigorous starburst during the last $< 40$\,Myr and
contains some of the most massive super star clusters in the nearby Universe. Given its starburst age and
 its star-formation rate, \eso\ is one of the most efficient nearby manufactures of neutron stars and black holes,
  hence providing an excellent laboratory for feedback studies.   }
   {We aim to use X-ray observations with the largest modern X-ray telescopes \xmm\ and \cxo\  to unveil the most luminous 
   accreting  neutron stars and black holes in \eso.
    }
   {We compared X-ray images and spectra with  integral field spectroscopic 
   observations in the optical  to constrain the nature of strong X-ray emitters.  
     }
   {X-ray observations uncover three ultraluminous X-ray sources (ULXs) in \eso. 
    The brightest among them,  ESO\,338~X-1, has X-ray luminosity in excess of 
   $10^{40}$\,erg\,s$^{-1}$. We speculate that \eso\,X-1 is powered by accretion 
   on an intermediate-mass ($\msim 300\,M_\odot$) black hole.  {\nw We show that 
   X-ray radiation from ULXs and {\new hot superbubbles} strongly contributes 
   to He\,{\sc ii} ionization and general stellar feedback in this template starburst galaxy.}}
   {}


\authorrunning{Oskinova et\,al.}
\titlerunning{Candidate IMBH and a population of ULXs in the starburst \eso\,-\,IG04}

   \maketitle
%

\section{Introduction}

Awareness of the important contribution from    
high-mass X-ray binaries (HMXBs) to the overall stellar feedback is 
growing quickly \citep[e.g.][]{Madau2004, Mirbel2011,Just2012,Prestwich2015, Madau2017}.   
Recently, \citet{Sazonov2018} showed that HMXB feedback is likely dominated 
by ultraluminous X-ray sources (ULXs), which are defined as off-nucleus point sources with 
X-ray luminosity $L_{\rm X}\msim 10^{40}$\,erg\,s$^{-1}$.  
In a star forming galaxy in the local 
Universe, there are about $\sim 0.4\,-\,1.2$ ULXs
per unit of star formation rate 
\citep[SFR $M_\odot\,{\rm yr}^{-1}$;][]{Map2010,SK2017}.
Despite being rare, ULXs in a dwarf galaxy significantly 
photoionize helium in the interstellar medium (ISM), hence  
reducing its opacity for the extreme UV and soft X-rays. This leads to a
longer mean free path of energetic photons, assisting the ISM heating, 
and in turn easing the driving of galactic outflows, 
and facilitating the escape of LyC and Ly$\alpha$ photons.       

It is becoming increasingly clear that the majority of ULXs are powered by super-critical 
accretion onto neutron stars (NSs) and stellar mass black holes  
\citep[BHs; see a recent 
review by ][]{kaaret2017}. {\new The super-critically accreting NS might reach X-ray 
luminosities above $\msim 10^{41}$\,erg\,s$^{-1}$ \citep{Israel2017}, but 
in the ULXs with $L_{\rm X} > 10^{40}$\,erg\,s$^{-1}$, accretion onto an 
intermediate-mass black hole (IMBH) is generally suspected \citep{Farrell2009}}.

One of the closest (37.5\,Mpc) luminous blue compact dwarf galaxies, \eso, provides 
an excellent laboratory to better understand and disentangle the various constituencies 
of stellar feedback. In optical and UV this galaxy shares a number of properties with faint 
Lyman break galaxies with weak but positive Ly$\alpha$ emission. \citet{Bik2018} presented 
a comprehensive study of \eso\ and its circumgalactic medium based  on  integral field 
spectroscopic observations in optical obtained with the 
the Multi Unit Spectroscopic Explorer (MUSE) instrument  at  the Very Large Telescope (VLT); for a 
complete overview of this galaxy we refer the interested readers to this work and 
references therein.  \eso\ has been undergoing a vigorous starburst for the last 
$\sim 40$\,Myr \citep{Ostlin98,Ostlin03} and contains a large population of young massive 
star clusters; among them 
are clusters displaying strong Wolf-Rayet (WR) spectral features. The most remarkable among them is cluster~23, the most luminous, massive 
($\sim 10^7\,M_\odot$), and young ($< 10$\,Myr) super-star cluster (SSC) detected in the 
nearby galaxies \citep{Ostlin07}. \citet{Bik2018} observed an extended ionized halo around \eso, 
SSCs immersed in a large superbubble filled by shocked gas, and a galactic scale wind. Indirect evidence suggests that \eso\ leaks ionizing 
LyC radiation  \citep{Leitet2013}. The total mass of \eso\ ($4\times 10^9\,M_\odot$), 
its current star formation rate, $3.2\,M_\odot\,{\rm yr}^{-1}$ \citep{Ost2001}, and 
the age of its massive clusters make it one of the most active nearby  producers of 
NSs and BHs.      

Here we report \xmm\ and \cxo\ X-ray observations 
of \eso. 
In Sect.~2 we present the X-ray observations and X-ray source 
identifications, while the discussion on the potential IMBH candidate 
and the contribution of ULXs to stellar feedback in this galaxy is 
presented in Sect.\,3. A summary is provided in Sect.\,4. 
  
\section{X-ray observations and source identification}

\begin{figure}[t]
\centering
\includegraphics[width=0.9\columnwidth]{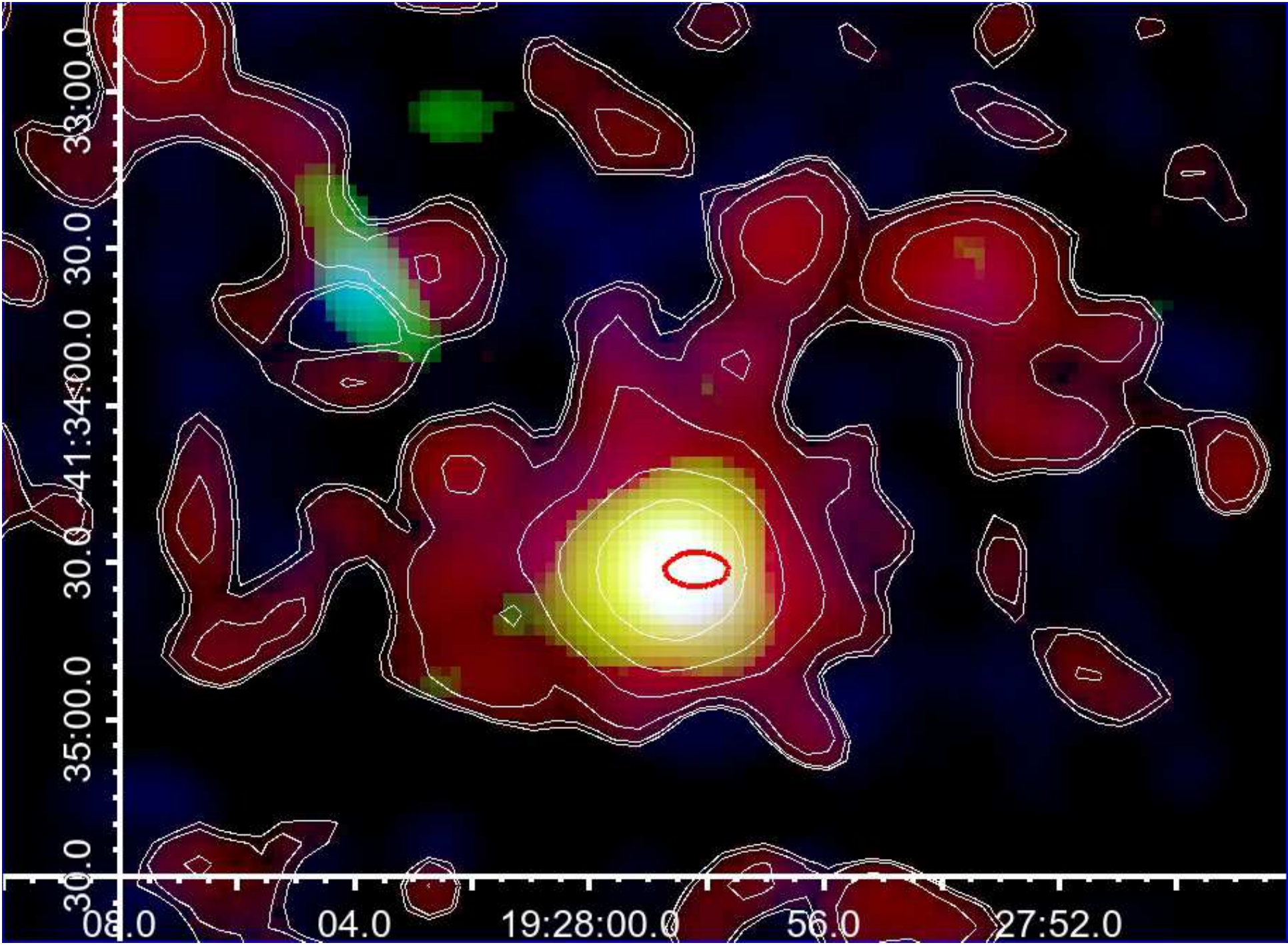}
\caption{Combined smoothed \xmm\  EPIC PN image of \eso\ in soft (0.2-1.0\,keV, red), 
medium (1.0-4.5\,keV, green) and hard (4.5-12.0\,keV, blue) energy bands with 
{\new the logarithmic-scale contours (white), tracing X-ray emission in the soft band.}   
The image size is $4'\times 3'$ and north is up and east to the left. The red 
ellipse ($16"\times 10"$) approximately marks the position and UV extent of the \eso\ galaxy.} 
\label{fig:ximg}
\end{figure}

\subsection{\xmm}
\label{sec:xmm}
        
\xmm\ observed \eso\ on 2016-04-10 (ObsID 0780790201; PI H.\,Ot\'{\i}--Floranes). 
We reprocessed these data using the \xmm\  Science Analysis 
System (SAS)\footnote{\url{www.cosmos.esa.int/web/xmm-newton/what-is-sas}}. 
The standard analysis steps were performed to obtain the X-ray images, spectra, and 
light curves. After rejecting high-background time intervals, the useful exposure time 
was 15\,ks for the EPIC PN and 23\,ks for the EPIC MOS cameras.

\begin{figure}[t]
\centering
\includegraphics[width=0.9\columnwidth]{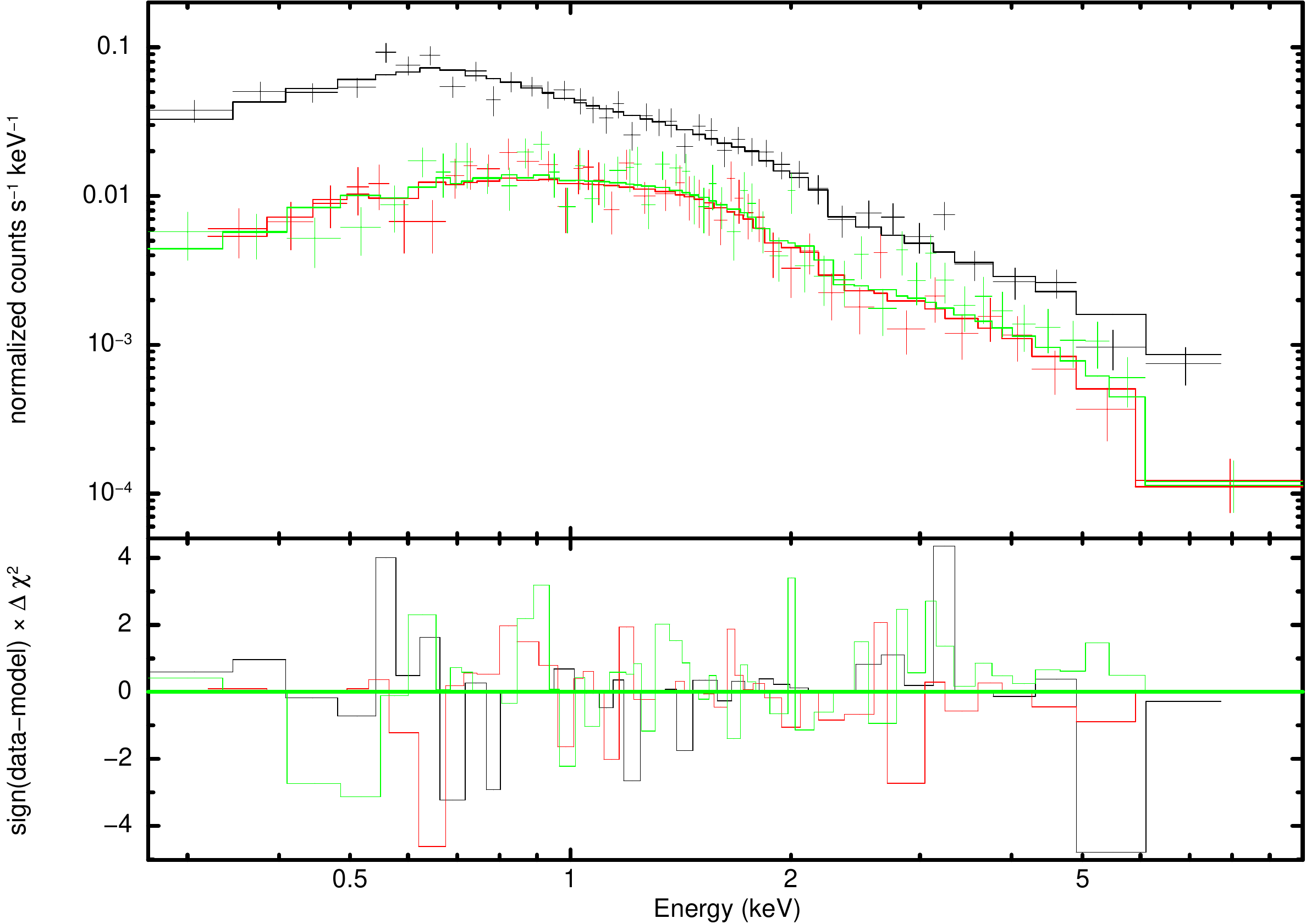}
\caption{\xmm\ EPIC PN,  MOS1, and MOS2
spectra of  3XMM J192758.7-413430  (black, red, and green curves, respectively)  
with error bars corresponding to 3$\sigma$ and  
the best-fit combined thermal plasma ({\it apec}) and power-law model 
(solid lines). The model parameters are shown in Table 1.  
The lower panel shows residuals between the data and the best-fit model.} 
\label{fig:xspec}
\end{figure}

An X-ray image of the sky around \eso\ is shown in Fig.\,\ref{fig:ximg}. A bright 
X-ray source is clearly detected at the location of \eso. This source, 3XMM J192758.7-413430, 
can be found in the ``\xmm\ Serendipitous Source Catalogue 3XMM-DR8'' at  
$\rm{RA}\,(2000)=19^{\rm{h}} 27^{\rm{m}} 58\fs75$,
$\rm{Dec.}\,(2000)=-41\degr 34\arcmin 30\farcs 8$ with an error in position $1\farcs 5$.
The observed source flux is $(3.2 \pm 0.1)\times 10^{-13}$\,\flux. 
No foreground or background objects coinciding with the X-ray source are
seen in the VLT MUSE and  Hubble Space Telescope ({\em HST}) 
images \citep{Bik2018}. We conclude 
that 3XMM J192758.7-413430 resides in the \eso\ galaxy. In this case its  
absorption-corrected X-ray luminosity in 0.2-12.0\,keV band is 
$8\times 10^{40}$\,erg\,s$^{-1}$.

\begin{table}
\begin{center}
\caption[ ]{Parameters of the best-fit spectral models
to the observed  \xmm\ EPIC spectra of 3XMM\,J192758.7-413430.
\cxo\ observations (see Sect. 2.2) show that the dominant 
contribution to the 3XMM\,J192758.7-413430 X-ray light is provided 
by the IMBH candidate  \eso~X-1. The  fitted models are the thermal plasma 
{\it apec} model for the metallicity $0.2\,Z_\odot$ combined  with a power law 
with the photon index of power law $\Gamma$, and corrected for 
ISM absorption {\it tbabs}. Fluxes and luminosity are in the 0.2-12.0\,keV band, {\new 
and fluxes are listed for the individual model components as well as for the full model.}}.
\begin{tabular}[]{lc}
\hline
\hline
Model parameter & Best-fit value \\
\hline
\multicolumn{2}{c}{\it tbabs(powerlaw+apec)} \\
\hline
$N_{\mathrm H}$ [$10^{21}$ cm$^{-2}$]   &$1.5\pm 0.3$  \\
$kT$ [keV]                                                    &$0.22\pm0.02$   \\
$EM$ [$10^{63}$ cm$^{-3}$]                       & $2\pm 1$   \\
$\Gamma$     &   $1.85 \pm 0.06$ \\                               
$K$ [photon\,keV$^{-1}$\,cm$^{-2}$\,s$^{-1}$ at 1\,keV] &  $(5.5\pm 0.4)\times 10^{-5}$ \\                          
reduced $\chi^2$ for 134 d.o.f. &  $0.88$ \\
{\new Unabsorbed} $F_{\rm X}^{\rm thermal}$ [\flux] & $7\times 10^{-14}$ \\
{\new Unabsorbed} $F_{\rm X}^{\rm powerlaw}$ [\flux] & $4\times 10^{-13}$ \\
\hline
\multicolumn{2}{c}{\it tbabs($N_{\rm H, 1})$*powerlaw+tbabs($N_{\rm H, 2}$)*apec} \\
\hline
$N_{\mathrm H, 1}$ [$10^{21}$ cm$^{-2}$]    & $0.6\pm 0.3$    \\
$kT$ [keV]                                                         &$0.24\pm0.02$  \\
$EM$ [$10^{63}$ cm$^{-3}$]                           & $2\pm 1$   \\
$N_{\mathrm H, 2}$ [$10^{21}$ cm$^{-2}$]    & $2.7\pm 1$    \\
$\Gamma$                                                        & $1.96 \pm 0.1$ \\                               
$K$ [photon\,keV$^{-1}$\,cm$^{-2}$\,s$^{-1}$ at 1\,keV] &  $(6.3\pm 0.7)\times 10^{-5}$ \\                          
Reduced $\chi^2$ for 133 d.o.f. &  $0.87$ \\
{\new Unabsorbed} $F_{\rm X}^{\rm thermal}$ [\flux] & $7\times 10^{-14}$ \\
{\new Unabsorbed} $F_{\rm X}^{\rm powerlaw}$ [\flux] & $4\times 10^{-13}$ \\
\hline
Model $F_{\rm X}$ [\flux] & $ 3\times 10^{-13} $ \\
Luminosity $L_{\rm X}$ [erg\,s$^{-1}$] & $8\times 10^{40} $\\
\hline
\hline
\end{tabular}
\label{tab:par}
\end{center}
\end{table}

Figure\,\ref{fig:ximg} shows indications of extended soft 
X-ray emission surrounding \eso, as indeed might be expected given the presence 
of a shocked superbubble and galactic-scale outflows in \eso\ \citep{Bik2018}.

\begin{figure*}[t!]
\centering
\includegraphics[width=1.99\columnwidth]{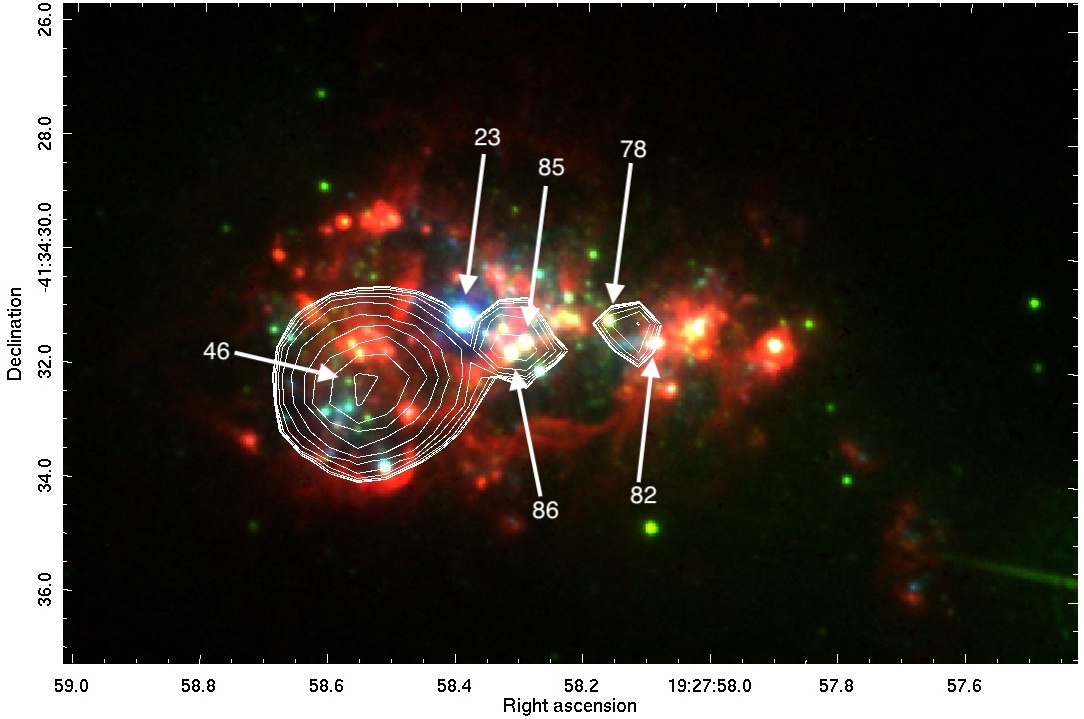}
\caption{{\em HST} colour composite image of \eso. The blue shows the image in 
F140LP UV filter, highlighting the youngest star clusters, green is the F550M filter, and red is the 
FR656N images centred on the H$\alpha$ emission line, highlighting the distribution of the ionized gas. 
Overplotted in white are  {\nw logarithmically spaced contours} tracing the smoothed  \cxo\ X-ray image. 
Three detected X-ray sources, X-1, X-2, and X-3, are aligned from left to right. The numbered 
arrows indicate the positions of star clusters that might be associated with  the X-ray sources 
(see discussion in Sect.~2.3 and Table\,2.) } 
\label{fig:multilam}
\end{figure*}

Indirect support for the presence of hot thermal plasma in \eso\ 
comes from the analysis of low-resolution \xmm\ EPIC spectra
(Fig.\,\ref{fig:xspec}). The spectra were analysed using standard X-ray 
spectral fitting software {\em xspec} \citep{xspec1996}. The fitting statistics 
improve when in addition to a single power-law model, a  $\sim 2$\,MK thermal 
plasma component is included in a spectral mode  {\new (reduced $\chi^2=0.886$ 
compared to reduced $\chi^2=0.977$ for a simple power-law model). The 
F-test  indicates (F-value $\approx 7$ and probability 0.001) that including an 
additional thermal plasma model component is justified.  Using
a black-body spectral component
instead   
does not improve fitting statistics compared to a single power-law model  
(reduced $\chi^2=0.969$,  F-value $\approx 0.6$ and probability 0.5).   
We speculate that  soft optically thin thermal plasma  
X-ray emission is associated with the superbubble and galactic wind.   } 
     
The power-law component has a spectral index $\Gamma\approx 1.9$
which is similar to the brightest ULXs and is 
well explained by accretion onto a BH \citep{Sutton2012}. 

Interestingly, spectral fits of similar quality could also be 
achieved with a purely thermal two-temperature model. In this case, 
a second temperature component with $kT= 5.3\pm 0.6$\,keV is recovered, 
which could indicate the presence of a hot accretion disk. {\new To compare with
various accretion-disk models  we tried the {\em diskpbb} 
model \citep{Kubota2005} combined with the {\em apec} model. A good-quality fit 
(reduced $\chi^2\approx 0.9$) is achieved with the temperature at inner disk radius, 
$kT_{\rm in}=3.3\pm 1.5$\,keV and the exponent of the radial dependence of the disk 
temperature $p=0.53\pm 0.02$ \citep[i.e.\  similar to other ULX spectra; see ][]{Pintore2018}.

The foreground Galactic absorption to the galaxy is 
$N_{\rm H}\approx 5\times 10^{20}$\,cm$^2$  \citep{Hayes2005}, that is,\
significantly lower than derived from the X-ray spectra fitting.  
\citet{Bik2018} determined the extinction map of the the galaxy and found 
that within the galaxy $E(B-V)$ varies between $\approx0.05$ and $0.18$\,mag. 
The largest local absorption in \eso\ could in principle explain the relatively 
high neutral hydrogen column density  derived from the X-ray spectral analysis.

However, the high column density of absorbers could also be intrinsic to the ULX
itself. To explore this possibility we tested a combined power law plus optically thin 
thermal plasma model, where each spectral component has its own absorbing column.  
The fitting results are shown in Table\,1. Indeed, in this case the 
thermal plasma could be well described as suffering the foreground absorption only, 
while the power-law component is significantly more strongly  absorbed.}

The standard timing analysis procedures were employed to search for a
period between 0.2\,s to 100\,s in the barycentre and background-corrected 
X-ray light curves of 3XMM\,J192758.7-413430. No period with sufficiently high 
statistical significance was uncovered.   

\subsection{\cxo} 

\begin{table*}
\begin{center}
\caption[ ]{Parameters$^{\rm A}$ of X-ray sources
detected by \cxo\ in the \eso\ galaxy. 
\cxo\ 99\%\ limit on positional accuracy is $1\farcs4$. Count rates, fluxes (observed), 
and luminosities (corrected for the ISM absorption) are in 0.5-7.0\,keV band.}
\begin{tabular}[]{ccccccc}
\hline
\hline
\eso\ & RA & Dec & Rate~~(90\%\ confidence) & $L_{\rm X}$    &nearest cluster & separation \\ 
         &      &          &  [$10^{-3}$\,s$^{-1}$]           &  [erg\,s$^{-1}$] & \multicolumn{2}{c}{\citep{Bik2018}}  \\ \hline 
X-1 & 291.99390 & -41.57566 & 13.10~~(9.9--17.1) &  $\sim 4\times 10^{40}$ &  46  & $0\farcs31$ \\ 
X-2 & 291.99293 & -41.57542 & ~2.21~~(1--4) & $\sim 1\times 10^{40}$ &   85/86 & 
$0\farcs19$/$0\farcs21$ \\
X-3 & 291.99219 & -41.57538 & ~1.43~~(0.5 --3) &  $\sim 3\times 10^{39}$     & 78/82 & $0\farcs 4$/$0\farcs 5$ \\
\hline
\hline
\end{tabular}
\label{tab:src}
\tablefoot{$^{\rm A}$ based on the output from the {\em wavedetect},   
{\em fluximage}, {\new and {\em srcflux}} tasks which are the part of CIAO package} 
\end{center}
\end{table*}

The angular resolution of the \xmm\ telescope ($\sim 11\arcsec$) 
is surpassed by that of the \cxo\ X-ray telescope ($0\farcs5$). 
\cxo\ observed \eso\ for 
3\,ks on 19 Nov 2018 (ObsID 21866, DDT observations). After standard data-reduction steps, 
the source detection was performed using the {\em wavdetect} tool, which accounts for the 
the point-spread function (PSF). The tool is a part of the \cxo\ 
Interactive Analysis of Observations data analysis package 
CIAO\footnote{\url{cxc.harvard.edu/ciao}}. 

Among the ten sources detected in the whole ACIS-S field of view, three are 
located within the optical and UV extent of the \eso\ galaxy (Fig.\,3). We denote them 
\eso~X-1, X-2, and X-3 in order of decreasing X-ray luminosity as observed on  19 Nov 2018 (Table\,2).
All three  sources are in the ULX luminosity range. The X-ray luminosity 
of the brightest source, \eso~X-1, is $L_{\rm X} (0.5-7\,{\rm keV})\approx 4\times 10^{40}$\,erg\,s$^{1}$. 
Assuming standard accretion, this corresponds to the Eddington luminosity of a $300\,M_\odot$ BH.         

Fewer than approximately $35$ counts were detected from \eso~X-1 (and even less from the other two sources). 
This is not sufficient for measuring X-ray spectra. However, the hardness ratio is consistent with the 
spectrum measured by \xmm.

No significant time variability was noticed within the 50\,min duration of the \cxo\ observations.
Interestingly, the PSF fitting shows that the ``size'' of \eso~X-1 exceeds the size of the PSF 
by a factor of about two. This might  be an indication that the source is extended, for example\ 
immersed in a diffuse X-ray-emitting nebula as is sometimes observed around 
other ULXs \citep{Soria2010, Webb2017}. 

{\new Keeping in mind the huge difference in the sensitivities and angular resolution 
between \xmm\ and \cxo\ exposures, it appears that within the errors the source 
fluxes did not change between 2016 and 2018 epochs.}    

\subsection{Potential optical counterparts of X-ray sources}

Figure\,\ref{fig:multilam} shows an overlay of the optical ({\em HST}) 
images \citep[from ][]{Ostlin09} tracing both the gas and the stars as well as the X-ray  (\cxo) contours of \eso. 
The 99\%\ limit on  the positional accuracy of \cxo\  is $1\farcs4$. The optical {\em HST} images are astrometically 
aligned to the MUSE datacube using isolated star clusters \citep[see ][]{Bik2018}.
The accuracy of the MUSE astrometry \citep[calibrated to the USNO UCAC4 catalogue,][]{Zacharias2013} 
is estimated to be around $0\farcs5$ or better. Unfortunately, the \cxo\ exposure was too short to detect any optical foreground stars  
in X-rays seen by the {\em HST} or MUSE  -- this 
would have allowed better astrometrical  alignment of X-ray, optical and UV images. At the 
moment, it is difficult to claim a confident identification of UV and optical counterparts of X-ray sources. 
However, complementing the {\em HST} images with the MUSE integral field spectroscopy in the 
optical is very useful to obtain initial clues as to the nature of the ULXs in \eso.         

The {\em HST} data have a completeness limit of about 50\%\ for a $5\times 10^3 M_\odot$ cluster
{\new at 50\,Myr and a $10^4 M_\odot$ cluster at 100\,Myr \citep{Ostlin03}}, and therefore massive 
clusters associated with the detected ULXs 
should be seen in the {\em HST} images. 
The distance between the enigmatic most massive 
cluster~23 and the X-ray sources X-1 and X-2 is 2\arcsec\ and 1\arcsec, respectively. This is comparable to 
the positional error of the X-ray sources. Hence we cannot exclude 
the association between cluster 23 and \eso~X-1 or X-2. 

Nearest to the centroid of the X-ray position of X-1 is the cluster~46
\citep[see the nomenclature in ][]{Bik2018} located at   
$19^{\rm{h}} 27^{\rm{m}} 58\fs56$,
$-41\degr 34\arcmin 32\farcs 25$. The separation between 
\eso~X-1 and cluster~46 is $0\farcs31$ ($\sim 55$\,pc) which is smaller 
than the \cxo\ angular resolution. 
The age of the cluster (10\,Myr) and its mass ($6\times 10^{4}\,M_\odot$) 
make it a plausible host of a massive donor star capable of   
supplying  a sufficiently high accretion rate 
to maintain the nearly Eddington X-ray luminosity of a $\msim 300\,M_\odot$ BH. 

Interestingly, as can be seen in Fig.\,\ref{fig:multilam}, there is a pronounced
arc-like structure to the south of the X-ray-emitting region containing \eso~X-1.
Future observations should show whether this is an unrelated ISM structure, an outer wall of 
a superbubble, or a bow shock produced by\ a jet powered by \eso~X-1 for example.   

Several massive young clusters
are found within the positional uncertainty of \eso~X-2. 
The closest coincidence is with two massive young clusters: 
cluster~86  ($19^{\rm{h}}27^{\rm{m}}58\fs308$, $-41\degr 34\arcmin 31\farcs72$) and 
cluster~85 ($19^{\rm{h}}27^{\rm{m}}58\fs286$, $-41\degr 34\arcmin 31\farcs53$). 
Both these clusters are $\sim 6$\,Myr old and have mass of $\sim 4\times 10^{5}\,M_\odot$.
MUSE optical spectroscopy of 85/86 clusters shows a hint of a broad He\,{\sc ii} line, 
suggestive of WR stars, which would be strongly expected given the cluster ages and 
masses. 

Source \eso~X-3 also has some young massive clusters within its $1\farcs5$ error circle: 
clusters 78 ($19^{\rm{h}}27^{\rm{m}}58\fs156$, $-41\degr 34\arcmin 31\farcs12$, 
10 Myrs, $ 5\times 10^5$ M$_\odot$) and 82 ($19^{\rm{h}}27^{\rm{m}}58\fs085$, $-41\degr 34\arcmin 31\farcs49$, 4\,Myr, 
$10^5$ M$_\odot$). It is tempting to suggest that \eso~X-2  and \eso~X-3 might be WR+BH binaries. 



\section{Discussion}

\subsection{\eso~X-1 -- an IMBH candidate}

The X-ray luminosity of \eso\,X-1 is in the upper range of the ULXs. In 
 a study of 63 ULXs only two were found to be more luminous than 
$10^{40}$\,erg\,s$^{-1}$ \citep{Wolter2018}. \citet{Earnshaw2019} presented a catalogue of non-nuclear 
X-ray sources in nearby galaxies. Among 384 ULXs in this catalogue, only approximately ten sources in the 40\,Mpc volume have X-ray luminosities in excess of 
$4\times 10^{40}$\,erg\,s$^{-1}$. \citet{Zolotukhin2016} presented a search for 
hyperluminous X-ray sources in the \xmm\ source catalogue and found 
about 15 candidates. These latter authors emphasised that only joint optical and X-ray imaging 
and spectroscopy can confirm their nature. 

The X-ray spectrum of \eso~X-1, a power law with $\Gamma\approx 1.9$, appears to 
be consistent with other IMBH candidates. Furthermore,  \eso~X-1 is located in a metal-poor galaxy undergoing a vigorous starburst over the last $\sim 40$\,Myr, with 
a large number of SSCs with ages of $\sim 10$\,Myr. At these ages the 
most massive stars are already collapsed, while there are still a large number
of massive donors at the WR and red supergiant (RSG) evolutionary stages 
 when their mass loss 
is most prodigious \citep[e.g.][]{PZ1996, Osk2005}.

{\new A causality link between young dense massive star clusters and progenitors of IMBHs
has long since been predicted \citep{Begelman1978}.  IMBHs  are expected 
to be associated with massive star clusters older than 5\,Myr \citep{PZ2002}. Indeed, detailed 
simulations show that IMBHs 
could form  at the centres of young, compact stellar clusters  \citep{Gur2006}.  The cluster core 
collapses relatively quickly  ($\lsim 5$\,Myr).  A runaway sequence of stellar mergers 
can lead to the formation of a very massive 
star ($M_\ast\msim 400\,M_\odot$)   \citep{Freitag2006}. 
Although the final fate of the resulting object is difficult to predict because of uncertainties in 
stellar evolution, among different potential outcomes is its direct collapse to an IMBH. Recently, \citet{Bel2019}  
performed  high-resolution cosmological simulations focusing on dwarf galaxies at zero-redshft. 
These latter authors found that larger dwarf galaxies are more likely to host massive black holes  than those of lower 
mass. Interestingly, in their simulations, about half of the IMBHs in dwarfs are located within a few kiloparsecs 
of the galaxy centre. Low-metallicity  (low-$Z$)  conditions (but higher than the primordial) are especially 
favourable for the  formation of IMBHs \citep[][]{Map2010}. 

So far, IMBHs have not been ambiguously identified \citep{Mezcua2017}.   {\nw  Close to 
the ESO\,243-49 galaxy is the strong IMBH candidate, HLX-1, with a BH of $>10\,000\,M_\odot$. The 
galaxy has modest ongoing star formation in the bulge of only $\sim 0.01\,M_\odot$\,yr$^{-1}$, and 
the observations rule out any young SSC close to HLX-1. This unique object 
might be a nucleus of a satellite dwarf galaxy  \citep{Soria2017}.}

The discovery of NS accretors in  ULXs strongly restricted their diagnostic potential for the 
IMBHs searches but did not rule out the existence of these latter objects. 
Nothing we know so far precludes  \eso~X-1 from being  a supercritical accreting NS 
or a BH of a few tens of solar mass   with a favorite 
beaming orientation. However, given that large 
theoretical and modeling studies predict that young star clusters with properties well 
sampled by those in  \eso\  provide conditions needed for IMBH formation, we 
must consider the possibility that \eso~X-1 hosts an IMBH. 
Furthermore, future observations showing that all ULXs in \eso\  are  
conventional sources would  strongly question our current understanding of 
massive young cluster evolution and dynamics at low-$Z$.   

Observations dictate that  ULXs are more common in low-metallicity galaxies \citep{Prestwich2013}, 
as indeed  predicted theoretically for the ULXs powered by accretion onto black holes \citep{Map2010}. } 
We suggest that the numbers of ULXs containing a NS  should have much weaker dependence on the 
host galaxy metallicity compared on the numbers of ULXs powered by the accretion 
onto BHs, {\new as also seen in the simulations of the stellar population evolution \citep{Wik2018}.} 
For the BHs in the upper mass range among those detected by the gravitational 
wave observatories, the ULX-range luminosities could be achieved at certain 
binary evolutionary stages without the need to invoke supercritical accretion, 
especially when the donor is a WR-type star \citep{Marchant2017, Hainich2018}. 
Given the high star formation rate in the nuclear region of \eso, a large number of 
stellar BHs are expected to have formed in the past.  The detection of  merging episodes of $30$--$40\,M_\odot$ 
BHs by their gravitational wave flashes shows that there are mechanisms by which 
quite massive BHs could form in massive star-forming regions. {\new Therefore,  ULXs powered 
by accretion onto  $20$--$60\,M_\odot$  BHs  could also be present in \eso.  An alternative scenario would 
be the formation of an IMBH during a previous star-formation episode in this galaxy. This IMBH might have 
acquired a donor star only recently.}

The apparent location of \eso~X-1 slightly offset from the most massive clusters in \eso\ is similar to  
other  bright ULXs. For example,\ \citet{PF2013} showed that ULXs in the Antennae galaxies are not
located in the star clusters, and  suggested that these ULXs are massive X-ray binaries that have 
been ejected from clusters by encounters involving several bodies. {\new \citet{Egorov2017} found 
kinematic evidence of the  escape of ULX HoII\,X-1 from a star cluster. }
In general, as reviewed by \citet{kaaret2017}, in spiral and moderately star-forming dwarf galaxies 
the ULXs are associated with loose clusters with masses of a few times 10$^3\,M_\odot$ and
ages of $\msim 10$\,Myr, or are located in non-star-forming regions.
{\nw 
In large starburst galaxies, bright ULXs are sometimes found near SSCs, 
for example\ the IMBH candidate M\,82~X-1 with an estimated mass of  $M_{\rm BH}\sim 1000\,M_\odot$  
\citep{Kaaret2001, Jang2018}. } \citet{vdH2013} proposed a scenario where accreting 
IMBHs (with $M_{\rm BH}=10^{2-3}\,M_\odot$)  could be companions in massive {\em runaway} binaries
formed in the dense cores of collapsed star clusters. 
Moreover, besides a stellar formation channel, other possible formation channels might be expected in a 
tidally interacting galaxy, such as \eso.
The improved astrometrical alignment of the {\em HST} and deeper \cxo\ images should help to clarify the
association between ULXs and star clusters in \eso.

Overall, given the present day observations and according to the criteria discussed in \citet{Sutton2012}, 
\eso\,X-1 appears to be an eligible accreting 
IMBH candidate.

\subsection{Contribution of ULXs to stellar feedback in \eso}

Using the advantages of integral field spectroscopy, \citet{Bik2018} derived 
ionization maps of \eso\ and its circumgalactic matter. 
They found  
that the central area of \eso\ is highly ionized, showing very high ratios in 
[O\,{\sc iii}]/[S\,{\sc ii}] and [O\,{\sc iii}]/H$\alpha$ (see their Fig.\,7 and corresponding discussion in their Sect.
4.4). The ionization 
potential of O$^+$ is $I_{\rm p}>35.2$\,eV. The spatial extent of the area 
with very high [O\,{\sc iii}]/[S\,{\sc ii}] strongly overlaps with the area 
filled by the nebular He\,{\sc ii} emission, indicating very high ionization. 
{\new The total He\,{\sc ii} emission, including both broad and narrow emission 
line components, is $\approx 1.1\times 10^{-14}$\,erg\,s$^{-1}$\,cm$^{-2}$, which 
corresponds to $L_{\rm HeII}\approx 2\times 10^{39}$\,erg\,s$^{-1}$.}
Interestingly, the [O\,{\sc iii}]/[S\,{\sc ii}] ratio is higher for lower optical 
depths of LyC photons \citep{Pel2012}. Our X-ray observations showed that the 
area of high ionization in \eso\ is affected by strong X-ray ionizing field from
the ULXs and likely the diffuse hot plasma.      
Existing X-ray observations detect only the brightest X-ray sources, while missing 
a large population of the fainter ones; for example,\ according to 
the cluster wind models, the cluster~23 alone should drive a hot cluster wind 
with $L_{\rm X}\msim 10^{37}$\,erg\,s$^{-1}$ \citep{CC85, Osk2005}, which is below the sensitivity of 
current \cxo\ observations. 

{\new  To estimate how many He\,{\sc ii} ionizing photons are produced by X-ray 
sources we need to know their flux  in the 0.0544\,--\,0.25\,keV 
energy band.  For this purpose we used the  {\em diskpbb+apec}  spectral model (see Sect.\,\ref{sec:xmm}) 
which better accounts for the flattening of the ULX spectra at low energies. 
The unabsorbed flux is $F_{\rm 0.0544-0.25\,keV}\approx  1 \times 10^{-13}$\,\,erg\,s$^{-1}$\,cm$^{-2}$, 
corresponding to $L_{\rm 0.0544-0.25\,keV}\approx 2\times 10^{40}$\,erg\,s$^{-1}$. 
Integrating the unabsorbed model spectrum over  the 0.0544-0.25\,keV energy interval  yields
the number of He\,{\sc ii} ionizing photons  $3\times 10^{50}$\,s$^{-1}$.  Assuming a branching ratio of
five to produce a photon in the He\,{\sc ii} $\lambda 4686$\,\AA\ line, the predicted luminosity 
of \eso\ in the He\,{\sc ii} line is  $\sim 1\times 10^{39}$\,erg\,s$^{-1}$ which is comparable to 
the observed one. 

X-rays from hot thermal plasma (described by the {\em apec} spectral component) dominate 
He\,{\sc ii} ionization. Neglecting the thermal plasma radiation and accounting only for the ULX spectrum 
(as described by the {\em diskpbb} model)  is responsible for only about 5\%\ of He\,{\sc ii} luminosity, 
$\sim 4\times 10^{37}$\,erg\,s$^{-1}$. In our model estimates we neglect local absorption.  Although
the metal-poor gas in \eso\ is especially transparent for soft X-rays,  our estimate of 
He\,{\sc ii} ionizing photons is likely an upper limit. 

Hot O and WR stars, which are abundant in this galaxy, further contribute to He\,{\sc ii} ionization. 
For example,\  a hot O star with $T_{\rm eff}=56$\,kK  at $Z=0.2Z_\odot$ produces 
$10^{47}$\,s$^{-1}$  He\,{\sc ii} ionizing photons, while an order of magnitude higher number of ionizing photons 
is produced by the hottest WN-type stars ($T_{\rm eff}=140$\,kK  at $Z=0.2Z_\odot$)\footnote{The rate of H, He\,{\sc ii}, and
He\,{\sc iii} ionizing photons could easily be retrieved from the public library of O and WR model 
atmospheres PoWR \url{www.astro.physik.uni-potsdam.de/PoWR} \citep{Todt2015, Hainich2019}}.
Given the large number of O and WR stars in \eso, and including soft X-rays, there 
is no need for exotic very massive or stripped stars to explain the ionization of \eso.  

Recently, \citet{Schaerer2019} concluded that both empirical data and theoretical models 
agree that high-mass X-ray binaries are the main source of nebular He\,{\sc ii} emission in 
low-$Z$ star-forming galaxies. At present, our study of \eso\  shows that the dominant agent 
of He\,{\sc ii} ionization is soft X-rays, likely emitted by shocked gas in superbubbles around 
star clusters.  However,  much deeper observations
 of this galaxy are required in order to make any firm conclusions.  

 \citet{Leb2017} suggested that X-ray heating quenches star formation on large scales in galaxies. 
 It seems that in galaxies like \eso, heating and ionization by X-rays could hinder sequential star formation. }

The number of ULXs detected in \eso\ is in a agreement with the expectations 
from the universal X-ray luminosity function which strongly depends on galactic 
SFR \citep[][]{Grimm2003}, and is consistent with other SFR 
indicators. We should note that \eso\ provides an important constraint for 
the calibrations of the dependence of the X-ray luminosity function on metallicity, 
which is presently not well known.

\section{Summary} 

\xmm\ and \cxo\ observations identified three ULXs in \eso, a galaxy which has been experiencing 
a strong episode of massive star formation for the last 40\,Myr. {\new Given the large number of 
young, $\sim 10$\,Myr-old SSCs in \eso\ and the large number of theoretical studies that independently 
predict the formation of IMBHs in SSCs of similar ages and metallicity, we propose that the 
brightest ULX  \eso~X-1, with a luminosity of $\msim 4\times 10^{40}$\,erg\,s$^{-1}$ is an IMBH 
candidate. }While the \cxo\ astrometric accuracy is not high enough to firmly associate the ULXs 
to some of the SSCs previously identified in \eso, the proximity of the ULXs to SSCs makes it 
very plausible that the ULXs are powered by the massive star donors. From \xmm\ observations, 
we have found some indications for the presence of extended soft X-ray emission surrounding \eso,  
which would be compatible with the presence of a shocked superbubble.  Finally, the hard ionizing 
flux produced by  ULXs, {\new together with the soft X-ray emission from the superbubbles} and the 
ionization produced by the abundant WR stars in this region, could explain the extended He\,{\sc ii} 
emission and the high level of ionization of the central area of \eso. We conclude, that the contribution of  ULXs 
to stellar feedback in the metal-poor starburst galaxy \eso\ is highly significant.

\begin{acknowledgements}
{\new The authors thank the referee for their very useful comments and suggestions which allowed to 
significantly improve the paper. The authors are grateful for the community response following 
the arXive  posting of the submitted manuscript --  numerous comments and suggestions 
were incorporated into the manuscript and strongly improved it. }  
The authors are grateful to \cxo\ observatory's Director for the DDT allocation.  
LMO  acknowledges financial support by the Deutsches Zentrum f\"ur Luft und Raumfahrt (DLR) 
grant FKZ 50 OR 1809, and partial support by the Russian Government Program of Competitive Growth of Kazan Federal University. 
JMMH is funded by Spanish State Research Agency grants
ESP2017-87676-C5-1-R and MDM-2017-0737 (Unidad de Excelencia 'Mar\'ia de
Maeztu' CAB).  A.A., A.B., M.H., and G.\"O. acknowledge the support of the Swedish Research Council, 
Vetenskapsr{\aa}det and the Swedish National Space Agency (SNSA).
      
\end{acknowledgements}


\end{document}